\documentclass[showpacs,preprintnumbers,amsmath,aps,amssymb]{revtex4}
\usepackage{graphicx}
\usepackage[english]{babel}
\usepackage{bm}
\begin{document}
\baselineskip 16pt
\title{Scalar potentials with Multi-scalar fields from quantum cosmology and supersymmetric quantum mechanics}
\author{J. Socorro}
\email{socorro@fisica.ugto.mx}
\author{Omar E. N\'u\~{n}ez}
\email{neophy@fisica.ugto.mx}

 \affiliation{Departamento de  F\'{\i}sica, DCeI, Universidad de Guanajuato-Campus Le\'on,
 C.P. 37150, Le\'on, Guanajuato, M\'exico}

\begin{abstract}

The Multi-scalar field cosmology of the anisotropic Bianchi type I
model is used in order to construct a family of potentials that are
the best suited to model the inflation phenomenon. We employ the
quantum potential approach to quantum mechanics due to Bohm in order
to solve the corresponding Wheeler-DeWitt equation; which in turn
enables us to restrict sensibly the aforementioned family of
potentials. Supersymmetric Quantum Mechanics (SUSYQM) is also
employed in order to constrain the superpotential function, at the
same time the tools from SUSY Quantum Mechanics are used to test the
family of potentials in order to infer which is the most convenient
for the inflation epoch. For completeness solutions to the wave
function of the universe are also presented.

\end{abstract}
 \maketitle

\section{Introduction}

The inflation phenomenon is one of the most accepted mechanism to
explain the early expansion of the universe, similar to that of
present day cosmic acceleration,  the quintessence scalar field
theory is the most commonly used in the literature \citep{santos,
andrewl, gomez, capone}, however if we add another quintessence
scalar field, i.e. a multi-scalar field theory it is possible to
explain the transition from late inflation to an early stage of
radiation epoch \cite{juanm}, in this sense the multi-scalar fields
cosmology is  a viable candidate to explain such phenomenon and for
that, an especific form of the potential for the scalar fields is
needed. The former is the main objetive of this work.

  In the present study we desire
to perform our investigation in the case of multi-scalar fields
cosmology, constructed using both quintessence fields, mantaining a
nonspecific potential form $\rm V(\phi,\sigma)$. There are many works in the literature
\citep{copeland2, quiros, bento, copeland3, coley} that deal with this type of problems, but
in a general way and not with a particular ansatz, but rather with one that only
considers dynamical systems. One special class of potentials used to
study this behaviour corresponds to the case of the exponential
potentials  for each field, where the corresponding energy density
 of a scalar field has the range of scaling behaviors \citep{andrew,ferreira}, i.e, it scales
exactly as a power of the scale factor like,  $\rho_\phi\propto
a^{-m}$, when the dominant component has an energy density which
scales in a similar way. There are some works where other type of
potentials are analyzed \cite{copeland}.

 How come that we claim that the analysis of general potentials using dynamical
systems was made considering particular structures of them,  in
other words, how can we introduce this mathematical structure within
a physical context?. We can answer this question, when the
Bohmian  and SUSYQM formalism are introduced, i.e, many of them can be
constructed using the Bohm formalism \citep{wssa,sodo,bohm} of the
quantum mechanics under the integral systems premise, which is known
as the quantum potential approach, furthermore with SUSYQM
we can narrow it down to the most suitable potential for inflation epoch.

This approach makes possible to identify trajectories associated with the wave function of the
universe \citep{wssa} when we choose the superpotential function as
the momenta associated to the coordinate field $\rm q^\mu$. This
investigation was undertaken within the framework of the
minisuperspace approximation of quantum theory but only for
 models with a finite number of degrees of freedom. Considering
the anisotropic Bianchi Class A cosmological models from canonical
quantum cosmology under determined conditions in the evolution of
our universe, and employing the Bohmian formalism, and in particular
the Bianchi type I we obtain a family of potentials that correspond
to the most probable to model the inflation phenomenon. In our
SUSYQM analysis, we found that the best candidate to model the
inflation phenomenon is an exponential potential, however in our
case this appeared as mixed in terms of the scalar fields and not as
linear combination of them.

This work is arranged as follows. In section \ref{model} we present
the corresponding Einstein Klein Gordon equation for the multi
scalar fields model. In section \ref{hap},  we introduced the
hamiltonian apparatus which is applied to Bianchi type I in order to
construct a master equation for all Bianchi Class A cosmological
models with barotropic perfect fluid and cosmological constant. In
section \ref{qap} we present the quantum scheme, where we use the
Bohmian formalism and show its mathematical structure, our
approach is also presented in a similar way, which is comparable to the
Wheeler-DeWitt equation when is expressed as an expansion in terms of $\hbar$,
but only up to second degree in $\hbar$.
Our treatment is applied to build the mathematical structure of multi scalar-field potentials using the integral systems formalism.
For completeness we present the quantum solutions to the Wheeler-DeWitt equation.
However it is important to emphasize that the quantum potential from Bohm formalism will work as a constraint equation
 which restricts the family of potentials found.
In section \ref{susyqm} we employ the tools of SUSYQM to narrow the family of potentials to the most suitable
 candidate for inflation, solutions to the wave function of the universe in the Grassmann variables are also presented.

\section{The model \label{model}}
We begin with the construction of the multi scalar field
cosmological paradigm, which requires the simultaneous consideration
of two fields, namely two canonical ($\sigma$,$\phi$), the action of
a universe with the constitution of such fields, the cosmological
term contribution and the matter as perfect fluid content, is
\begin{equation}
\rm {\cal L}=\sqrt{-g} \left( R-2\Lambda
+\frac{1}{2}g^{\mu\nu}\nabla_\mu \phi \nabla_\nu \phi
+\frac{1}{2}g^{\mu\nu}\nabla_\mu \sigma \nabla_\nu \sigma -
V(\phi,\sigma)\right)+ {\cal L}_{matter}, \label{lagra}
\end{equation}
and the corresponding field equations becomes
\begin{eqnarray}
  \rm G_{\alpha \beta}+ g_{\alpha \beta} \Lambda &=&\rm +\frac{1}{2}\left(\nabla_\alpha \phi \nabla_\beta \phi -\frac{1}{2}g_{\alpha \beta}
  g^{\mu \nu} \nabla_\mu \phi \nabla_\nu \phi \right)\nonumber\\
  &&+
\frac{1}{2}\left(\nabla_\alpha \sigma \nabla_\beta \sigma
-\frac{1}{2}g_{\alpha \beta} g^{\mu \nu}
\nabla_\mu \sigma \nabla_\nu \sigma \right) \nonumber \\
&& \rm -\frac{1}{2}g_{\alpha \beta} \, V(\phi,\sigma) -8\pi G T_{\alpha \beta}, \label{munu}\\
\rm g^{\mu\nu} \phi_{,\mu\nu} -g^{\alpha \beta} \Gamma^\nu_{\alpha
\beta} \nabla_\nu \phi - \frac{\partial V}{\partial \phi}&=&\rm
0,\qquad
\Leftrightarrow \quad \Box \phi -\frac{\partial V}{\partial \phi}=0 \nonumber\\
\rm g^{\mu\nu} \sigma_{,\mu\nu} -g^{\alpha \beta} \Gamma^\nu_{\alpha
\beta} \nabla_\nu \sigma - \frac{\partial V}{\partial \sigma}&=&\rm
0,\qquad
\Leftrightarrow \quad \Box \sigma -\frac{\partial V}{\partial \sigma}=0,\nonumber\\
 \rm T^{\mu\nu}_{\,\,;\mu}&=&\rm 0,\quad with \quad T_{\mu\nu}= Pg_{\mu\nu}+(P+\rho)u_\mu u_\nu, \label{energy-momentum}
\end{eqnarray}
here $\rho$ is the energy density, P the pressure, and   $\rm u_{\mu
}$ the velocity, satisfying  that $\rm u_\mu u^\mu=-1$.


\section{Hamiltonian approach \label{hap}}
Let us recall here the canonical formulation in the ADM formalism of
the diagonal Bianchi Class A cosmological models. The metric has the
form
\begin{equation}
\rm ds^2= -N(t)dt^2 + e^{2\Omega(t)}\, (e^{2\beta(t)})_{ij}\, \omega^i
\, \omega^j, \label {met}
\end{equation}
where $\rm \beta_{ij}(t)$ is a 3x3 diagonal matrix, $\rm \beta_{ij}=
diag(\beta_++ \sqrt{3} \beta_-,\beta_+- \sqrt{3} \beta_-,
-2\beta_+)$, $\Omega(t)$ is a scalar and $\rm \omega^i$ are
one-forms that  characterize  each cosmological Bianchi type model,
and obey the form $\rm d\omega^i= \frac{1}{2} C^i_{jk} \omega^j \wedge
\omega^k$, and $\rm C^i_{jk}$  are structure constants of the
corresponding model.

The corresponding metric of the Bianchi type I in Misner's
parametrization  has the following form
\begin{equation}
\rm ds^2_I=-N^2 dt^2 +e^{2\Omega +2\beta_++2\sqrt{3}\beta_-}dx^2
+e^{2\Omega +2\beta_+ -2\sqrt{3}\beta_-}dy^2 +e^{2\Omega
-4\beta_+}dz^2,\label{metricI}
\end{equation}
where the anisotropic radii are
$$\rm R_1=e^{\Omega +\beta_+ +\sqrt{3}\beta_-}, \qquad R_2=e^{\Omega +\beta_+
-\sqrt{3}\beta_-}, \qquad  R_3=e^{\Omega -2\beta_+ }.$$

 We use the Bianchi type I  cosmological model as toy
model to apply the formalism.
The lagrangian density (\ref{lagra})
for the Bianchi type I is written as
  (where the overdot denotes time derivative),
\begin{equation}
\rm {\cal L}_I = \rm e^{3\Omega}\left[6 \frac{\dot \Omega^2}{N}-6
\frac{\dot \beta_+^2}{N}-6 \frac{\dot \beta_-^2}{N} - 6 \frac{\dot
\varphi^2}{N}-6 \frac{\dot \varsigma^2}{N}+ N\left(
V(\varphi,\varsigma)+2\Lambda+16\pi G   \rho   \right)  \right],
\label{lagra-i}
\end{equation}
the fields were re-scaled as $\phi=\sqrt{12}\varphi,
\sigma=\sqrt{12}\varsigma$ for simplicity in the calculations.

 The momenta are defined as $\Pi_{q^i}=\rm \frac{\partial {\cal L} }{\partial \dot q^i}$, where
 $\rm q^i=(\beta_\pm,\Omega, \varphi,\varsigma)$ are the coordinates
fields.
\begin{eqnarray}
\rm \Pi_\Omega  &=&\rm \frac{\partial {\cal L}}{\partial \dot
\Omega} = \frac{12e^{3\Omega} \dot\Omega}{N}, \qquad \to
\dot \Omega = \frac{N\Pi_\Omega}{12}e^{-3\Omega} \nonumber\\
\rm \Pi_{\pm}  & =&\rm \frac{\partial {\cal L}}{\partial \dot
\beta_\pm} =-12\frac{e^{3\Omega} \dot\beta_\pm}{N}, \qquad  \to
\dot{{\beta_\pm}}  = -\frac{N\Pi_\pm}{12}  e^{-3\Omega}\\
\rm \Pi_{\varphi}  & =&\rm \frac{\partial {\cal L}}{\partial \dot
\varphi} =-12\frac{e^{3\Omega} \dot\varphi}{N}, \qquad  \to
\dot{{\varphi}}  = -\frac{N\Pi_\varphi}{12}  e^{-3\Omega}\nonumber\\
\rm \Pi_{\varsigma} &=&\rm \frac{\partial {\cal L}}{\partial \dot
\varsigma} = -12\frac{e^{3\Omega} \dot\varsigma}{N}, \qquad  \to
\dot{{\varsigma}}  = -\frac{N\Pi_\varsigma}{12}
e^{-3\Omega}.\nonumber \label{momenta}
\end{eqnarray}
Writing (\ref{lagra-i}) in canonical form,
 $\rm {\cal L}_{canonical}=\Pi_q \dot q -N{\cal H}_I$, when we perform the variation of this canonical
lagrangian with respect to N, $\frac{\delta {\cal
L}_{canonical}}{\delta N} =0$, implying the constraint ${\cal
H}_I=0$. In our model the only constraint corresponds to Hamiltonian
density, which is weakly zero. Now, substituting the energy density
for the barotropic fluid, we can find the Hamiltonian density ${\cal
H}_I$ in the usual way {\small{\begin{equation} \rm {\cal H}_I =
\frac{e^{-3\Omega}}{24}\left[\Pi_\Omega^2-\Pi_\varsigma^2-\Pi_+^2-\Pi_-^2-\Pi_\varphi^2
    -e^{6\Omega}\left\{24V(\varphi,\varsigma)+48\left( \Lambda+8\pi GM_\gamma
    e^{-3(\gamma+1)\Omega}\right)\right\}
\right]. \label{hami-i}
\end{equation}}}

where  we have used the covariant derivative of
(\ref{energy-momentum}), obtaining the relation
$$
3\dot \Omega \rho +3\dot \Omega p+\dot \rho=0,
$$
whose solution becomes
\begin{equation}
\rm \rho =M_\gamma e^{-3 \left( 1+\gamma \right)\Omega}.
 \label{density}
\end{equation}
where $M_\gamma$ is an integration constant.

 Considering the inflationary phenomenon $\gamma=-1$, the Hamiltonian density is
\begin{equation}
\rm {\cal H}_{I} = \rm
\frac{e^{-3\Omega}}{24}\left[\Pi_\Omega^2-\Pi_\varsigma^2-\Pi_+^2-\Pi_-^2-\Pi_\varphi^2
    - e^{6\Omega}\left\{24 V(\varphi,\varsigma)+
    \lambda_{eff}\right\}
\right], \label{hami-ix-1}
\end{equation}
where $\rm \lambda_{eff}=48(\Lambda+8\pi G M_{_{-1}})$.

\section{quantum approach\label{qap}}
On the Wheeler-DeWitt (WDW) equation there are a lot of papers
dealing with different problems, for example in \cite{Gibbons}, they
asked the question of what a typical wave function for the universe
is. In Ref. \cite{Zhi} there appears an excellent summary of a paper
on quantum cosmology where the problem of how the universe emerged
from big bang singularity can no longer be neglected in the GUT
epoch. On the other hand, the best candidates for quantum solutions
become those that have a damping behavior with respect to the scale
factor, since these allow to obtain good classical solutions when using
the WKB approximation for any scenario in the evolution of our
universe \cite{HH,H}. Our goal in this paper deals with the problem
to build the appropriate scalar potential for the inflationary
scenario.

 The Wheeler-DeWitt equation   for
this model is acquired by replacing
 $\rm \Pi_{q^\mu}=-i\hbar \partial_{q^\mu}$ in (\ref {hami-i}).
 The factor $\rm e^{-3\Omega}$ may be factor ordered with $\rm \hat \Pi_\Omega$ in many ways. Hartle and
Hawking \citep{HH} have suggested what might be called a
semi-general factor ordering, which in this case would order $\rm
e^{-3\Omega} \hat \Pi^2_\Omega$ as
\begin{eqnarray}
\rm - e^{-(3- Q)\Omega}\, \partial_\Omega e^{-Q\Omega}
\partial_\Omega&=&\rm - e^{-3\Omega}\, \partial^2_\Omega + Q\,
e^{-3\Omega} \partial_\Omega, \label {hh}
\end{eqnarray}
where  Q is any real constant that measure the ambiguity in the
factor ordering for the variable $\Omega$. In the following we will assume such factor ordering for the Wheeler-DeWitt equation,
which becomes
\begin{equation}
\rm \hbar^2 \Box \Psi+ \hbar^2 Q\frac{\partial \Psi}{\partial
\Omega}- e^{6\Omega}U(\beta_\pm,\varphi,\varsigma,\Lambda)\Psi=0,
\label{wdwmod}
\end{equation}
where $\rm \Box=-\frac{\partial^2}{\partial
\Omega^2}+\frac{\partial^2}{\partial
\varsigma^2}+\frac{\partial^2}{\partial \varphi^2}
+\frac{\partial^2}{\partial \beta_-^2} +\frac{\partial^2}{\partial
\beta_+^2}$ is the d'Alambertian in the coordinates
$q^\mu=(\Omega,\beta_\pm,\varsigma,\varphi)$ and the potential is $\rm
U= \left(48\Lambda +24V(\varphi,\varsigma) + 384 \pi G M_\gamma
e^{-3(\gamma-1)} \right)$. In the next section we introduce the main
idea of the Bohm formalism, and why we choose the phase in the wave
function to be real and not imaginary.

\subsection{Mathematical structure in the Bohm formalism}
In this section we will explain how the quantum potential approach or as
is also known, the Bohm formalism \cite{bohm}, works in the context of
quantum cosmology. For the cases that will be object of our
investigation in the sections to come, it is sufficient to consider
the simplest model, for which the whole quantum dynamics resides in
this single equation,
\begin{equation}
{\cal H}\psi=\left(g^{\mu \nu} \nabla_\mu \nabla_\nu -
V(q^\mu)\right)\psi=0, \label{basic}
\end{equation}
where the metric must be $\rm q^\mu$ dependent. The $\psi$ is called
the wave function of the universe, and we consider that $\psi$ has
the following traditional decomposition
\begin{equation}
\rm \psi=R(q^\mu)\, e^{\frac{i}{\hbar}S(q^\mu)}, \label{an}
\end{equation}
with $R$ and $S$ as real functions. Inserting (\ref{an}) into (\ref{basic}),
we obtain two equations corresponding to the real and imaginary
parts respectively, which are
\begin{eqnarray}
\rm \Box R - R \left[\frac{1}{\hbar^2} \left( \nabla S \right)^2 + V
\right]&=&\rm \Box R - R \left[H(S)\right]= 0,\label{real}\\
\rm 2\nabla R \cdot \nabla S + R \Box S&=&\rm 0, \label{imaginary}
\end{eqnarray}
when we consider the problem of factor ordering, usually in
cosmological problems, as indicated in the beginning of this
section, equation (\ref{hh}) must be included as a linear term of
$Q\frac{\partial \psi}{\partial q}$, where Q is a real parameter
that measures the ambiguity in this factor ordering. So, the
equations (\ref{real},\ref{imaginary}) are written as
\begin{eqnarray}
\rm \Box R +Q\frac{\partial R}{\partial q} -R\left[\frac{1}{\hbar^2}
\left( \nabla S \right)^2 + V
\right]&=&\rm 0,\label{rreal}\\
\rm 2\nabla R \cdot \nabla S + R \Box S+R \frac{\partial S}{\partial
q} &=&\rm 0, \label{iimaginary}
\end{eqnarray}
where $q$ is a single field coordinate.

We assume that the wave function $\psi$ is a solution of equation
(\ref{basic}), thus, this equation is equally satisfied.
Considering the Hamiton-Jacobi analysis, we can identify the
equation (\ref{rreal}) as the most important equation of this
treatment, because with this equation we can derive the time dependence, and then, it serves as the evolutionary equation in this
formalism. Following the Hamilton-Jacobi procedure, the $\Pi_q$ momenta is related to the superpotential function
S, as $\rm \Pi_{q^\mu}=\frac{\partial S}{\partial q^\mu}$, which are
related with the classical momenta (\ref{momenta}) written in the
previous section, hence,
\begin{equation}
\rm \frac{dq^\mu}{dt}=g^{\mu\nu} \frac{\delta H(S)}{\delta
\frac{\partial S}{\partial q^\nu}},
\end{equation}
which defines the trajectory $\rm q^\mu$ in terms of the phase of
the wave function S. We substitute this equation into (\ref{rreal}), and we find (using $\dot q^\mu=\frac{dq^\mu}{dt}$ and $\hbar=1$),
\begin{equation}
\rm  \left[\Box R +Q\frac{\partial R}{\partial q}\right] = R \left[
g_{\mu\nu} \dot q^\mu \dot q^\nu + V \right] .\label{realreal}
\end{equation}
Therefore we see that the quantum evolution differs from the
classical one only by the presence of a quantum potential term
$  \left[\Box R +Q\frac{\partial R}{\partial q}\right]$
on the left-hand side of the equation of motion. Since we assume
that the wave function is known, the quantum potential term is also
known.

In the next subsection we choose $\psi=W e^{-S/\hbar}$ as an ansatz
for the wave function. It was first remarked by Kodama
\cite{kodama1,kodama2} that the solutions to the Wheeler-DeWitt
(WDW) equation in the formulation of Arnowitt-Deser and Misner (ADM)
and the Ashtekar formulation (in the connection representation) are
related by $\psi_{ADM}=\psi_A e^{\pm i \Phi_A}$, where $\Psi_A$ is
the homogeneous specialization for the generating functional of the
canonical transformation between ADM variables to Ashtekar's ones
\cite{ashtekar}. This function was calculated explicitly for the
diagonal Bianchi type IX model  by Kodama, who also found
$\Psi_A=const$ as a solution, with $\Psi_A$ pure imaginary, for a
certain factor ordering. One expects a solution of the form $\psi=W
e^{\pm \Phi}$, where W is a constant, and $\Phi=i\Phi_A$. In fact
this type of solution has been found for the diagonal Bianchi Class
A cosmological models \cite{mory,obso}, but in some cases W is a
function, as we will see in our present study.

\subsection{Our treatment}
 Using the ansatz for the wavefunction
\begin{equation}
\rm \Psi=EXP\left[ \pm \frac{a_1}{\hbar} \beta_+ \pm \frac{a_2}{\hbar}
\beta_-\right]\Xi(\Omega,\varsigma,\varphi),\label{aniso}
\end{equation}
 the WDW equation
is read as
\begin{equation}
\rm \left[\hbar^2\Box + \hbar^2 Q\frac{\partial }{\partial \Omega}-
e^{6\Omega}U(\varphi,\varsigma,\lambda_{eff})+c^2\right]\Xi=0,
\label{wdwmod1}
\end{equation}
where $c^2= (a_1^2+a_2^2)$ and now $\Box$ is written in the reduced
coordinates $\ell^\mu=(\Omega,\varsigma,\varphi)$

We find that the WDW equation is solved when we choose an ansatz
similar to the one employed in the Bohmian formalism
 of quantum mechanics \citep{bohm}, so we
 make the following Ansatz for the wave function
\begin{equation}
\rm \Xi(\ell^\mu) = W(\ell^\mu) e^{-
\frac{S_\hbar}{\hbar}(\ell^\mu)}, \label{ans}
\end{equation}
 where $\rm S_\hbar(\ell^\mu)$ is known as the superpotential function, and W is the amplitude of probability that is
 employed in Bohmian
  formalism \citep{bohm}. Then (\ref {wdwmod1}) transforms into
\begin{equation}\rm
 \hbar^2 \left[{\Box \, W} - \frac{1}{\hbar}W {\Box \, S_\hbar} - \frac{2}{\hbar} {\nabla W}\cdot {\nabla
 S_\hbar}+\frac{1}{\hbar^2} W  \left(\nabla S_\hbar\right)^2\right]
 +\hbar^2 Q\left[ \frac{\partial W}{\partial \Omega} -  \frac{1}{\hbar}W \frac{\partial
S_\hbar}{\partial \Omega}\right] - {\cal U} W = 0, \label {mod}
\end{equation}
writing this equation as power in $\hbar$, we have
\begin{equation}
\rm \hbar^2\left[ {\Box \, W} + Q \frac{\partial W}{\partial
\Omega}\right] - \hbar \left[W {\Box \, S_\hbar} + 2 {\nabla W}\cdot
{\nabla  S_\hbar}+ Q W \frac{\partial S_\hbar}{\partial
\Omega}\right] + W\left[\left(\nabla S_\hbar \right)^2- {\cal U}
\right]=0.
\end{equation}
So, we can see that the contribution to quantum potential term
appears at $\hbar^2$ in the approximation to the Hamilton-Jacobi
like equation and the imaginary part corresponds at the $\hbar$ term in
this expansion.

The notation read as, $\rm \Box = G^{\mu
\nu}\frac{\partial^2}{\partial \ell^\mu
\partial \ell^\nu}$, $\rm {\nabla \, W}\cdot {\nabla \, \Phi}=G^{\mu
\nu} \frac{\partial W}{\partial \ell^\mu}\frac{\partial
\Phi}{\partial \ell^\nu}$, $\rm (\nabla)^2= G^{\mu
\nu}\frac{\partial }{\partial \ell^\mu}\frac{\partial }{\partial
\ell^\nu}= -(\frac{\partial}{\partial \Omega})^2+
(\frac{\partial}{\partial \varsigma})^2 +(\frac{\partial}{\partial
\varphi})^2 $, with $\rm G^{\mu \nu}= diag(-1,1,1)$, $\rm {\cal
U}=e^{6\Omega}U(\varphi,\varsigma,\lambda_{eff})-c^2$ is the
potential term for the cosmological model under consideration.

 Eq  (\ref{mod}) can be written as the following set of partial
differential equations
\begin{subequations}
\label{WDWa}
\begin{eqnarray}
\rm (\nabla S_\hbar)^2 - {\cal U} &=& 0, \label{hj} \\
\rm    \Box \, W + Q \frac{\partial W}{\partial \Omega} & = & 0 \label{cons}  \,\\
\rm   W \left( \Box S+ Q \frac{\partial S_\hbar}{\partial \Omega}
  \right) + 2 \nabla \, W \cdot \nabla \, S_\hbar &=& 0 \, , \label{wdwmo}  \, .
\end{eqnarray}
\end{subequations}
The first two equations correspond to the real part in a separated
way, also, the first equation is called the Einstein-Hamilton-Jacobi
equation (EHJ), and the third equation is the imaginary part, such
as the equations presented the in previous section (\ref{rreal},
\ref{iimaginary}).

 Following the references
\citep{wssa,sodo}, first, we shall choose to solve Eqs. (\ref{hj})
and (\ref{wdwmo}), whose solutions at the end will have to fulfill
Eq. (\ref{cons}), which will play the role of a constraint equation.

Taking the ansatz
\begin{equation}
\rm
S_\hbar(\Omega,\varsigma,\varphi)=\frac{e^{3\Omega}}{\mu}g(\varphi)
h(\varsigma)+c\left(b_1 \Omega + b_2 \Delta \varphi +b_3 \Delta
\varsigma \right), \label{super-po}
\end{equation}
where $\Delta \varphi=\varphi-\varphi_0$, $\Delta
\varsigma=\varsigma-\varsigma_0$ with $\varphi_0$ and $\varsigma_0$
as constant scalar fields, and $\rm b_i$  as arbitrary constants. Then, Eq
(\ref{hj}) is transformed as
\begin{eqnarray}
 && \rm  \frac{e^{6\Omega}}{\mu^2}\left[h^2\left(\frac{dg}{d\varphi}
\right)^2+g^2\left(\frac{dh}{d\varsigma} \right)^2-9 g^2 h^2-\mu^2
U(\varphi,\varsigma,\lambda_{eff})
\right]\nonumber\\
&&\rm +\frac{6ce^{3\Omega}}{\mu} \left[-b_1gh
+\frac{b_2}{3}h\frac{dg}{d\varphi}+\frac{b_3}{3}g\frac{dh}{d\varsigma}
\right] +c^2\left(-b_1^2+b_2^2+b_3^2+1 \right)=0. \label{ecuacion}
\end{eqnarray}

At this point we question ourselves how to solve this equation in
relation to the constant c, implying the behavior of the universe
with the anisotropic parameter $\beta_\pm$.
\begin{enumerate}
\item{}  When we consider this equation as an expansion in powers of $e^\Omega$, then  each term is null in a separated way,
but maintaining that the constant $\rm c\not=0$,
 \begin{eqnarray}
 \rm -b_1^2+b_3^2+b_2^2+1 &=&0,\label{con1}\\
\rm -b_1gh
+\frac{b_2}{3}h\frac{dg}{d\varphi}+\frac{b_3}{3}g\frac{dh}{d\varsigma}&=&0,
\label{con2}\\
\rm h^2\left(\frac{dg}{d\varphi}
\right)^2+g^2\left(\frac{dh}{d\varsigma} \right)^2-9g^2 h^2-\mu^2
U(\varphi,\varsigma,\lambda_{eff})&=&0,\label{con3}
\end{eqnarray}
The first equation have the constraint between the constants $\rm
b_1^2=1+b_2^2 + b_3^2$, and the second equation gives the possible
solution for the function g and h,
$$\rm g=g_0 e^{\frac{3\eta_1}{b_2}\Delta \varphi}, \qquad h=h_0 e^{\frac{3\eta_2}{b_3}\Delta
\varsigma},$$ with the constraint between the separation constants
$\eta_i$, $\eta_1+\eta_2=b_1$ and the corresponding scalar field
potential
$$\rm U(\varphi,\varsigma,\lambda_{eff})=U_0 e^{\lambda_1\Delta \varphi+ \lambda_2 \Delta \varsigma}
$$
with $\rm U_0=\frac{9}{\mu^2}\frac{g_0^2 h_0^2}{(b_2
b_3)^2}\left[\eta_1^2 b_3^2+\eta_2^2 b_2^2-(b_2b_3)^2 \right]$,
$\lambda_1=\frac{6\eta_1}{b_2}$ and $\lambda_2=\frac{6\eta_2}{b_3}$

Using the superpotencial function $\rm
S=\frac{e^{3\Omega}}{\mu}g(\varphi)h(\varsigma)+c\left(b_1 \Omega +
b_2\varphi + b_3 \varsigma \right)$, and the ansatz for the amplitud
of probability $\rm W=e^{u(\Omega)+v(\varphi)+z(\varsigma)}$, the
equation (\ref{wdwmo}) is written as
\begin{eqnarray}
&&\rm \frac{e^{3\Omega}}{\mu}\left[-3(3-Q)gh+h
\frac{d^2g}{d\varphi^2}+ g\frac{d^2h}{d\varsigma^2} -6gh
\frac{du}{d\Omega}+2h\frac{dg}{d\varphi}
\frac{dv}{d\varphi}+2g\frac{dz}{d\varsigma} \frac{dh}{d\varsigma}
\right] \nonumber\\
&& \rm + c\left( Qb_1-2b_1 \frac{du}{d\Omega}+2b_2
\frac{dv}{d\varphi}+2b_3 \frac{dz}{d\varsigma}\right)=0,
\label{wdwmo-1}
\end{eqnarray}
using again the expansion in powers of $e^\Omega$, we have the
solutions for the functions u,w and z as
\begin{equation}
\rm u=\left(\frac{\alpha_1}{2b_1}+\frac{Q}{2}\right) \Omega + u_0,
\qquad v=\frac{\alpha_2}{2b_2}\varphi + v_0, \qquad
z=\frac{\alpha_3}{2b_3} \varsigma + z_0, \nonumber
\end{equation}
where $\alpha_i$ are separation constants, satisfying the relation
$\rm \alpha_1=\alpha_2+\alpha_3$, and the constraint between the
constants
\begin{equation}
\rm \alpha_2 b_3^2\left(-b_2^2+b_1 \eta_1\right) + \alpha_3 b_2^2
\left(-b_3^2+b_1 \eta_2 \right) + 3b_1\left(\eta_1^2b_3^2+\eta_2^2
b_2^2 - b_2^2 b_3^2\right)=0. \nonumber
\end{equation}

Also, the equation (\ref{cons}) produces the constrain
\begin{equation}
 \rm -\alpha_1^2 b_2^2 b_3^2+\alpha_2^2b_1^2 b_3^2 + \alpha_3^2
b_1^2 b_2^2 + b_1^2 b_2^2 b_3^2 Q^2=0. \nonumber
\end{equation}
 Finally, the wave functions for this models becomes
$$\rm \Xi(\ell^\mu)=W_0\,Exp\left[\left(\frac{\alpha_1}{2b_1}+\frac{Q}{2}+\frac{cb_1}{\hbar}\right) \Omega+ \left(\frac{\alpha_2}{2b_2}+\frac{cb_2}{\hbar}\right)  \varphi
 +\left( \frac{\alpha_3}{2b_3}+\frac{cb_3}{\hbar}\right)
 \varsigma\right]\, Exp\left[ Exp\left[3\left(\omega + \frac{\eta_1}{b_2}\varphi + \frac{\eta_2}{b_3}\varsigma\right)\right]
 \right].
 $$

\item{} For the case c=0, we have the following.

The constants $\rm a_i$ are related as $\rm a_2=\pm i a_1$, hence
the wave function corresponding to the anisotropic behavior becomes
$\rm e^{\pm a_1 \beta_+ \pm i a_1\beta_-}$, i.e, one part goes as
oscillatory in the anisotropic parameter.
\end{enumerate}
Now we use the case c=0 to obtain the appropriate
potential fields in the inflation phenomenon.
\subsection{Mathematical structure of potential fields}
To solve the Hamilton-Jacobi equation (\ref{hj})
$$\rm -\left(\frac{\partial S}{\partial \Omega} \right)^2+\left(\frac{\partial S}{\partial \varsigma} \right)^2
+\left(\frac{\partial S}{\partial \varphi} \right)^2=e^{6\alpha}
U(\varphi,\varsigma,\lambda_{eff})$$
 we propose that the superpotential function has the form
\begin{equation}
\rm S=\frac{e^{3\Omega}}{\mu} g(\varphi) h(\varsigma), \label{sp}
\end{equation}
and the potential
\begin{equation} \rm U=g^2 h^2 \left[a_0 G(g) + b_0 H(h)\right], \label{potenti-u}
\end{equation}
where $\rm g(\varphi)$, $\rm h(\varsigma)$, $\rm G(g)$ and $\rm
H(h)$ are generic functions of the arguments, which will be
determined under this process. When we introduced the ansatz in
(\ref{hj}) we found the following master equations for the fields
$(\varphi,\varsigma)$, (here $\rm c_1=\mu a_0$ and $\rm
c_0=\mu b_0$)

 Then, by separation of variables we find the
following master equations for the scalar fields
\begin{subequations}
\label{fields}
\begin{eqnarray}
\rm d\varphi&=&\rm \pm \frac{dg}{g\sqrt{\ell^2+ c_1 G}}, \qquad with \quad \ell^2=\nu^2-\frac{9}{2}>0,\label{varsigma}\\
\rm d\varsigma&=&\rm \pm \frac{dh}{h\sqrt{p^2+ c_0 H}}, \qquad with
\quad p^2=\nu^2+\frac{9}{2},\label{varphi}
\end{eqnarray}
\end{subequations}
where $\nu$ is a constant of separation of variables.

For  particular choices of functions $\rm G$ and $\rm H$  we can
solve $\rm g(\varphi)$ and $\rm h(\varsigma)$ functions, and
then use them to obtain the potential term U from (\ref{potenti-u}).
Some examples are shown in the tables \ref{t:solutions}, \ref{potentials} and \ref{potentials2}, thereby, the superpotential $\rm
S(\Omega,\varphi)$ is known, and the possible multifields potentials
are shown in the tables  \ref{potentials} and \ref{potentials2}.

\begin{center}
    \begin{table}[h]
            \begin{tabular}{|c|c|c|c|c|}
            \hline
            $\rm  H(h)$ & $\rm h(\varsigma)$ &$\rm G(g)$& $\rm g(\varphi)$ for $\nu^2>\frac{9}{2} $ & $\rm g(\varphi)$ for $\nu^2<\frac{9}{2} $   \\ \hline
            $ 0 $ & $\rm h_0 e^{\pm p  \Delta \varsigma}$&  0& $\rm g_0 e^{\pm \ell  \Delta \varphi}$ & $\rm g_0 e^{\pm \ell  \Delta \varphi}$ \\ \hline
            $\rm H_0$ & $\rm h_0 e^{\pm \sqrt{p^2+c_0 \, H_0} \Delta \varsigma}$& $\rm  G_0$ &$\rm g_0 e^{\pm \sqrt{\ell^2+c_1 \,G_0} \Delta \varphi}$ & $\rm g_0 e^{\pm \sqrt{c_1 \,G_0 - \ell^2} \Delta \varphi}$ \\ \hline
            $\rm  H_0 h^{2} $ &$\rm h_0 csch\left[p \Delta \varsigma \right]  $&  $\rm  G_0 g^{2}  $
            & $\rm g_0 csch \left[\ell \Delta \varphi \right] $ & $\rm g_0 csc \left[\ell \Delta \varphi \right] $ \\ \hline
            $\rm  H_0 h^{-2} $ &$\rm h_0 sinh\left[p \Delta \varsigma \right]  $&  $\rm  G_0 g^{-2}  $
            & $\rm g_0 sinh \left[\ell \Delta \varphi \right] $ & $\rm g_0 sin \left[\ell \Delta \varphi \right] $\\ \hline
            $\rm H_0 h^{-n}$ ($\rm n \neq 2$) &  $\rm h_0\, \left[sinh^2\left(\frac{n p \Delta \varsigma }{2} \right)\right]^{1/n}$ & $\rm G_0 g^{-n}$ ($\rm n \neq 2$)&  $\rm g_0\, \left[sinh^2\left(\frac{n \ell \Delta \varphi}{2}\right)\right]^{1/n}$ &  $\rm g_0\, \left[sin^2\left(\frac{n \ell \Delta \varphi}{2}\right)\right]^{1/n}$ \\\hline
            $\rm  H_0\, \ln h$ & $\rm h_0 e^{u(\varsigma)}$, & $\rm  G_0\, \ln g$& $\rm e^{v(\varphi)}$ & $\rm e^{v(\varphi)}$ \\
            & $\rm u(\varsigma)=\left(\frac{p}{2}\Delta \varsigma \right)^2$&
            &$\rm v(\varphi)=\left(\frac{\ell}{2}\Delta \varphi \right)^2$ & $\rm v(\varphi)=\left(\frac{\ell}{2}\Delta \varphi \right)^2$ \\\hline
            $\rm H_0  (\ln h)^2$ & $\rm e^{r(\varsigma)} $ & $\rm G_0  (\ln g)^2$ &$\rm e^{\omega(\varphi)} $ & $\rm e^{\omega(\varphi)} $\\
            & $\rm r(\varsigma)=sinh(p  \Delta \varsigma)  $&
            &$\rm \omega(\varphi)=sinh(\ell\,  \Delta \varphi) $ & $\rm \omega(\varphi)=cosh(\ell\,  \Delta \varphi) $ \\  \hline
            \end{tabular}
             \caption{ \label{t:solutions} \emph{Some exact solutions to eqs.
(\ref{varsigma},\ref{varphi}), where n is any real number, $\rm G_0$
and $\rm H_0$ are an arbitrary constants. Both cases for $\rm g(\varphi)$ have been considered in relation to the constant $\ell^2$ }}
        \end{table}
    \end{center}

    \begin{center}
        \begin{table}[h]
             \begin{tabular}{|c|c|}
            \hline
            $\rm  U(\varphi,\varsigma)$  with $\nu^2>\frac{9}{2}$ & Relation between all constants  \\ \hline
            0 &  $\rm \ell^2 s^2+\mu_0^2 p^2 +\ell^2 p^2(-k^2+Q^2+\ell^2+p^2-4)=0$  \\ \hline
            $\rm U_0 e^{\pm 2[\sqrt{p^2+c_0H_0}\Delta \varsigma + \sqrt{\ell^2+c_1G_0}\Delta \varphi]}$ &
             $\rm (s-p^2-3k-9-c_1G_0)^2-( p^2+c_1G_0)(s-\ell^2+c_0H_0)^2 $\\
                 & $\rm  +(p^2+c_0H_0)( p^2+c_1G_0)(k^2-Q^2)=0$\\             \hline
            $\rm U_0 csch^4(\ell\Delta \varphi) csch^2(p \Delta \varsigma)+U_1\, csch^4(p \Delta \varsigma) csch^2(\ell\Delta \varphi)$&$\rm k^2+2(\ell^2+p^2+Q^2+s+\mu_{0}+4)=0,\,\,3\ell^4+2\ell^2\mu_0 +2\ell^2+\mu^2_0=0,\,\,\,$\\
              & $\rm 3p^4+2sp^2+2p^2+s^2=0$ \\ \hline
            $\rm U_0 sinh^2(\ell\Delta \varphi)+U_1\, sinh^2(p \Delta \varsigma)$&$\rm 3\ell^4-2\ell^2\mu_0-2\ell^2+\mu_0^2=0,\,\,Q^2-k^2 +2(s-p^2 -\ell^2+\mu_0)=0,\,\,\,$\\
              & $\rm 3p^4-2sp^2-2p^2+s^2=0$ \\ \hline
            {\small{$\rm b_0H_0 \left[\frac{c_1G_0}{p^2} sinh^2\left(\frac{n p\Delta \varphi}{2}\right)\right]^{\frac{2}{n}}
            \left[\frac{c_0H_0}{\ell^2} cosh^2\left(\frac{n}{2}\ell\Delta \varsigma\right)\right]^{\frac{2-n}{n}} +$}} &
            $\rm  \mbox{quantum constraint only allows $n=\pm 2$} $   \\
            {\small{$\rm a_0G_0 \left[\frac{c_1G_0}{p^2} sinh^2\left(\frac{n p\Delta \varphi}{2}
            \right)\right]^{\frac{2-n}{n}}
            \left[\frac{c_0H_0}{\ell^2} cosh^2\left(\frac{n}{2}\ell\Delta \varsigma\right)\right]^{\frac{2}{n}} $}} &$\rm  \mbox{it reduces to previous cases} $  \\ \hline
            $\rm e^{2u(\varsigma)+2v(\varphi)}\left[b_0H_0 u(\varsigma)+
            a_0G_0\,v(\varphi) \right]$ &$\rm  \mbox{quantum constraint is not satisfied} $\\\hline
                        $\rm e^{2r(\varsigma)+2\omega(\varphi)}\left[b_0H_0 r^2 +a_0G_0 \omega^2 \right]$&  $
                        \mbox{quantum constraint is not satisfied} $ \\ \hline
                      \end{tabular}
                      \caption{ \label{potentials} \emph{The corresponding multifield potentials
that emerge from quantum cosmology in direct relation with the table
(\ref{t:solutions}). We also present the relation between all the
constants that satisfy the eqn. (\ref{cons}). We can see that the
quantum constraint restricts the general potential of the fifth line to
remain in the state of $n=\pm 2$. The sixth and seventh lines indicate
that these potentials are not allowed. }}
        \end{table}
    \end{center}

The other cases correspond to $\rm
\nu^2<\frac{9}{2}$, thus (\ref{varsigma}) reads as
 $$\rm
d\varphi =\rm \pm \frac{dg}{g\sqrt{ c_1 G-\ell^2}}, \qquad with
\quad \ell^2=\nu^2-\frac{9}{2},$$ and we can repeat the same
procedure to find the new function $\rm g(\varphi)$, the function $\rm h(\varsigma)$ remains the same for this segment.
The results are shown in table \ref{potentials2}.

    \begin{center}
        \begin{table}[h]
             \begin{tabular}{|c|c|}
            \hline
            $\rm  U(\varphi,\varsigma)$  with $\nu^2<\frac{9}{2}$ & Relation between all constants  \\ \hline
            0 &  $\rm \ell^2 s^2+\mu_0^2 p^2 +\ell^2 p^2(-k^2+Q^2+\ell^2+p^2-4)=0$  \\ \hline
            $\rm U_0 e^{\pm 2[\sqrt{p^2+c_0H_0}\Delta \varsigma + \sqrt{c_1G_0 - \ell^2}\Delta \varphi]}$ &
             $\rm (c_1G_0-\ell^2)(c_0H_0+p^2)(c_1G_0+c_0H_0-k^2+Q^2-\ell^2+p^2-4)$\\
                 & $\rm  +s^2(c_1G_0-\ell^2)+\mu^2_0(c_0H_0+p^2)=0$\\             \hline
            $\rm U_0 csc^4(\ell\Delta \varphi) csch^2(p \Delta \varsigma)+U_1\, csch^4(p \Delta \varsigma) csc^2(\ell\Delta \varphi)$&$\rm k^2+2(p^2-Q^2-\ell^2+s+\mu_{0}+4)=0,\,\,3\ell^4-2\ell^2\mu_0 -2\ell^2+\mu^2_0=0,\,\,\,$\\
              & $\rm 3p^4+2sp^2+2p^2+s^2=0$ \\ \hline
            $\rm U_0 sin^2(\ell\Delta \varphi)+U_1\, sinh^2(p \Delta \varsigma)$&$\rm 3\ell^4+2\ell^2\mu_0+2\ell^2+\mu_0^2=0,\,\,Q^2-k^2 +2(s-p^2 +\ell^2+\mu_0)=0,\,\,\,$\\
              & $\rm 3p^4-2sp^2-2p^2+s^2=0$ \\ \hline
            {\small{$\rm U_0sin^{4/n-2}{(\frac{n}{2}\ell \Delta \varphi)} sinh^{4/n}{(\frac{n}{2} p \Delta \varsigma)}+$}} &
            $\rm  \mbox{quantum constraint only allows $n=\pm 2$} $   \\
            {\small{$\rm +U_1 sinh^{4/n-2}{(\frac{n}{2} p \Delta \varsigma)} sin^{4/n}{(\frac{n}{2} \ell \Delta \varphi)}$}} &$\rm  \mbox{it reduces to previous cases} $  \\ \hline
            $\rm e^{2u(\varsigma)+2v(\varphi)}\left[b_0H_0 u(\varsigma)+
            a_0G_0\,v(\varphi) \right]$ &$\rm  \mbox{quantum constraint is not satisfied} $\\\hline
                        $\rm e^{2r(\varsigma)+2\omega(\varphi)}\left[b_0H_0 r^2 +a_0G_0 \omega^2 \right]$&  $
                        \mbox{quantum constraint is not satisfied} $ \\ \hline
                      \end{tabular}
                      \caption{ \label{potentials2} \emph{The corresponding multifield potentials
that emerge from quantum cosmology in direct relation with the table
(\ref{t:solutions}) but using the last column for $\rm g(\varphi)$ corresponding to $\rm
\nu^2<\frac{9}{2}$. The relation between all
constants that satisfy the eqn (\ref{cons}) are also present.}}
        \end{table}
    \end{center}


To solve (\ref{wdwmo}) we assume that
 \begin{equation}\rm
W=e^{u(\Omega)+v(\varphi)+z(\varsigma)}, \label{ww}
\end{equation}
and introducing the corresponding superpotential function S
(\ref{sp}) into the equation (\ref{wdwmo}), it follows the equation
\begin{equation}
\rm 3(-3+Q)-6\frac{du}{d\Omega} +\frac{1}{g}
\frac{d^2g}{d\varphi^2}+\frac{2}{g}\frac{dg}{d\varphi}
\frac{dv}{d\varphi}+\frac{1}{h}\frac{d^2h}{d\varsigma^2}+\frac{2}{h}\frac{dz}{d\varsigma}\frac{dh}{d\varsigma}=0,
\label{solve}
\end{equation}
and using the method of separation of variables,
  we arrive to a set of ordinary differential equations for
the functions $u(\Omega)$, $\rm v(\varphi)$ and $z(\varsigma)$. However, this decomposition is not unique, as it depends
 on how we choose the constants in the equations.
\begin{eqnarray}
\rm 2\frac{d\eta}{d\Omega}-Q&=&\rm k, \\
\rm \frac{d^2g}{d\varphi^2} +2\frac{dg}{d\varphi} \frac {dv}{d\varphi}&=& \rm [-s+3(k+3)]g,\\
\rm \frac{d^2h}{d\varsigma^2}+2\frac{dh}{d\varsigma}
\frac{dz}{d\varsigma}&=& \rm sh,
\end{eqnarray}
whose solutions in the generic fields g and h are
\begin{eqnarray}
\rm u(\Omega)&=& \rm \frac{Q+k}{2} \Omega, \nonumber\\
\rm z(\varsigma)&=&\rm \frac{s}{2}\int
\frac{d\varsigma}{\partial_\varsigma(ln h)}-\frac{1}{2}\int
\frac{\frac{d^2h}{d\varsigma^2}}{\partial_\varsigma h}d\varsigma,
\nonumber\\
\rm v(\varphi)&=& \rm \left(-\frac{s}{2} +\frac{3k}{2}
+\frac{9}{2}\right)\int \frac{d\varphi}{\partial_\varphi(ln g)}
-\frac{1}{2}\int \frac{\frac{d^2g}{d\varphi^2}}{\partial_\varphi
g}d\varphi, \nonumber
\end{eqnarray}
then
\begin{equation}
\rm W=e^{ \frac{s}{2}\int
\left(\frac{d\varsigma}{\partial_\varsigma(ln
h)}-\frac{d\varphi}{\partial_\varphi(ln g)} \right)} e^{-\frac{1}{2}
\int \left(\frac{\frac{d^2h}{d\varsigma^2}}{\partial_\varsigma
h}d\varsigma+ \frac{\frac{d^2g}{d\varphi^2}}{\partial_\varphi
g}d\varphi \right)} e^{\frac{k}{2}\left(\Omega+3\int
\frac{d\varphi}{\partial_\varphi(ln g)}\right)}
e^{\frac{1}{2}\left(Q\Omega+9\int
\frac{d\varphi}{\partial_\varphi(ln g)}\right)}. \label{ww2}
\end{equation}

In a similar way, the constraint (\ref{cons}) can be written as
\begin{equation}
 \rm \partial^2_{\varphi} v + \left(\partial_{\varphi} v \right)^2+\partial^2_{\varsigma}z
 + \left( \partial_{\varsigma}z \right)^2+\frac{Q^2-k^2}{4} = 0 \, , \label{wdwho2}
\end{equation}
or in other words (here $\rm \mu_0=-s+3(3+\kappa)$)
\begin{eqnarray}
&& \rm -2\frac{\partial^3_\varsigma h}{\partial_\varsigma
h}-2\frac{\partial^3_\varphi g}{\partial_\varphi g}
-2(s+1)h\frac{\partial^2_\varsigma h}{(\partial_\varsigma h)^2}
-2(\mu_0+1) g\frac{\partial^2_\varphi g}{(\partial_\varphi
g)^2}+3\left(\frac{\partial^2_\varsigma h}{\partial_\varsigma
h}\right)^2+3\left(\frac{\partial^2_\varphi g}{\partial_\varphi
g}\right)^2+s^2\left(\frac{ h}{\partial_\varsigma
h}\right)^2\nonumber\\
&&\rm +\mu^2_0 \left(\frac{ g}{\partial_\varphi
g}\right)^2+2s+2\mu_0+Q^2-k^2=0. \nonumber
\end{eqnarray}

Therefore, under canonical quantization we were able to determine a family of potentials that are
the most probable to characterize the inflation phenomenon in the evolution of our universe.

Now, we use the tools of SUSY Quantum Mechanics to test this family
of potential to infer which is more convenient for inflation era.

\section{Supersymmetric Quantum Mechanics for multi-scalars fields\label{susyqm}}

We use Witten's idea \cite{witten}, to find the supersymmetric
supercharges operators $\rm Q$ and $\rm \bar Q$ that produce a
superHamiltonian $\rm H_{ss}$, where the WDW equation can be
obtained as the bosonic sector of this super-Hamiltonian in the
superspace, i.e, when all fermionic fields are set equal to zero
(classical limit). It could be pointed that it may not be
 justified to use an effective bosonic action and the supersymmetrization,
 arising from a fundamental supersymmetric theory, due that the fermionic fields that
 appear under this approach, could not to be the same in both formalism.
 However, we can consider this approach as a toy model in such a way that the new
 fundamental fields effects arise from  the fundamental theory. The correct steps to
 supersymmetrize a bosonic Lagrangian, are to consider the true supersymmetry transformation
 in the sense of superfield scheme into the bosonic Lagrangian,  then the fermionic terms
 will emerge in a natural way \cite{tkach,donets}.

In this approach, the  supercharges for the 3D case  read as
\begin{equation}
\rm Q = \psi^\mu \left[-\hbar \partial_{q^\mu} + \frac{\partial
S}{\partial q^\mu} \right], \qquad \rm \bar Q = \bar \psi^\nu
\left[-\hbar \partial_{q^\nu} - \frac{\partial S}{\partial q^\nu}
\right], \label{supercharge2}
\end{equation}
where the $\rm S$ corresponds to equations (\ref{sp}),
 and the following algebra for the variables  $\psi^\mu$ and $\bar \psi^\nu$,
\begin{equation}\rm
\left\{ \psi^\mu ,\bar \psi^\nu \right \} = \eta^{\mu\nu}, \qquad
\left\{ \psi^\mu, \psi^\nu \right \} = 0, \qquad \left\{ \bar
\psi^\mu,\bar \psi^\nu \right \} =0.
\end{equation}
Using the representation $\rm \psi^\nu=\theta^\nu$ y $\rm \bar
\psi^\mu=\eta^{\mu\nu} \frac{\partial}{\partial \theta^\nu} $, one can find the superspace Hamiltonian in the form
\begin{equation}\rm
H_{ss}=\left \{ Q, \bar Q \right \} = {\cal H}_0 + \hbar
\frac{\partial^2 S}{\partial q^\mu \partial q^\nu} \left[ \psi^\mu,
\bar \psi^\nu \right] , \label{superhamiltonian}
\end{equation}
where $\rm {\cal H}_0=\square - U(q^\mu)$ is the standard WDW
equation, $\square$ is the 3D
 d'Alambertian in the $\rm q^{\mu}$ coordinates with $\rm \eta_{\mu \nu}=diag(-1,1,1)$,
 $\rm \{\, ,\, \}$  represent the anticommutator, and $\rm [\, ,\,]$ the commutator.

 The supercharges  $\rm Q, \bar Q$ and the  super-Hamiltonian satisfy the following
algebra
\begin{equation}\rm
 \left\{ Q,\bar Q\right \} = H_{ss}, \qquad
\left [H_{ss},Q \right]= \left[ H_{ss}, \bar Q \right ]=0,
\end{equation}

In this approach the supersymmetric physical states are selected by
the constraints
\begin{equation} \rm
 Q \, \Psi =0,\qquad {\bar Q}\, \Psi = 0 ,
\label{physical}
\end{equation}
this simplifies the problem of finding supersymmetric ground states
because the energy is known a priori and also the factorization of
$\rm H_{ss}|\Psi>=0$ into (\ref{physical}), often provides a simple
first-order equation for the ground state wave function. The
simplicity of this factorization is related to the solubility of
certain bosonic hamiltonians. It is well know that the existence of
normalizable solutions of the system (\ref{physical}) means that
supersymmetry is quantum mechanically unbroken.

The wave function has the  following decomposition in the  3D
Grassmann variables representation
\begin{equation}
\rm \Psi= {\cal A}_+ + {\cal B}_\nu \theta^\nu +
\frac{1}{2}\epsilon_{\mu\nu\lambda} \, {\cal C}^\lambda \,
\theta^\mu \, \theta^\nu + {\cal A}_- \, \theta^0 \, \theta^1 \,
\theta^2, \label{wavefun}
\end{equation}
$\mu, \nu, \lambda$ running over $0,1,2.$

Introducing the ansatz
\begin{equation}\rm
{\cal B}_{\nu} = \frac{\partial f_+ (q^{\nu})}{\partial q^{\nu}}\,
e^{\frac{S(q)}{\hbar}},
\end{equation}
into Eqs. (\ref{physical}) and (\ref{wavefun}), where the function
$\rm S$ is the superpotential function obtained as a solution for the
Einstein-Hamilton-Jacobi equation, Eq. (\ref{hj}) leads to the master
equation for the auxiliary function $\rm f_+$

\begin{equation}\rm
\hbar \Box f_+ + 2  \eta^{\mu\nu}
 \frac{\partial S}{\partial q^{\mu}}  \frac{\partial
f_+}{\partial q^{\nu}}=0. \label{master1}
\end{equation}
In addition, it is possible to show that $\rm \frac{1}{2}
\epsilon_{\mu\nu\lambda} {\cal C}^\lambda \theta^\alpha \theta^\mu
\theta^\nu= {\cal C}^\alpha \theta^0 \theta^1 \theta^2$ and
employing the ansatz

\begin{equation}\rm
{\cal C}^{\mu} = \eta^{\mu\nu} \frac{\partial f_-}{\partial q^{\nu}}
e^{-\frac{S(q)}{\hbar}},
\end{equation}
we obtain the second master equation in the form
\begin{equation}\rm
\hbar \Box f_- - 2  \eta^{\mu\nu}
 \frac{\partial S}{\partial q^{\mu}}  \frac{\partial
f_-}{\partial q^{\nu}}=0.
 \label{master2}
\end{equation}
Thus, Eqs. (\ref{master1}) and (\ref{master2}) can be written as
\begin{equation}
\rm \hbar \square \,f_\pm \pm 2 \eta^{\mu \nu} \frac{\partial
S}{\partial q^\mu} \frac{\partial f_\pm}{\partial q^\nu}=0.
\label{master3}
\end{equation}

The equations for the other functions ${\cal A}_\pm$ reads as
\begin{equation}\rm
\left[ \hbar\frac{\partial}{\partial q^{\mu}} \mp \frac{\partial
S}{\partial q^{\mu}}\right] {\cal A}_{\pm} =0.
\end{equation}
whose solutions  are
\begin{equation}\rm
{\cal A}_{\pm} = {a_0}_{\pm}\, e^{\pm \frac{1}{\hbar} S}, \label{As}
\end{equation}
where $\rm {a_0}_{\pm}$ are integration constants.

\subsection{Superquantum solution}
To solve Eq. (\ref{master3}) it is necessary to know the
superpotential function $\rm S(q^\mu)$. Once $\rm f_\pm$ are
obtained, all the bosonic component that appear in the Grassmann
expansion of the wave function (\ref{wavefun}) are determined.

The trivial solution  $\rm f_\pm =constants$, yields that the only
contributions to wave function are ${\cal A}_{\pm} $, which is
in agreement with the WKB proposal.

We want to write (\ref{master3}) as an homogeneous linear equation
of second degree
\begin{equation}
\rm \square W_\pm = W_\pm g(q^{\mu}),
\end{equation}
by introducing the ansatz into (\ref{master3})
\begin{equation}
\rm f_{\pm} = W_\pm (q^{\mu}) e^{\pm \phi(q^{\mu})/\hbar},
\end{equation}
we obtaining a  wave-like equation
\begin{equation}
\rm \square W_\pm \pm W_\pm \square S - W_\pm (\nabla S)^2 =0,
\end{equation}
it can be represented as
\begin{equation}
\rm \square W_\pm = g(q^{\mu}) W_\pm,
\end{equation}
where  $\rm g(q^{\mu}) = (\nabla S)^2 \mp \square S$. To solve it,
we propose a wave-like ansatz
\begin{equation}
\rm W_\pm = \beta_{\pm} e^{\mp s},
\end{equation}
which give us a condition on the s function
 $$\rm [(\nabla s)^2 \mp\square s] = [(\nabla S)^2 \mp\square S]$$
and if we propose that
\begin{equation}\rm
s = S \mp h(q^{\mu}),\qquad {\mbox with} \qquad h(q^{\mu})= m_\mu
q^{\mu},
\end{equation}
where $\rm m_\mu=(m_0,m_1,m_2)$ is a no null vector (the trivial
case in where $\rm h(q^{\mu})= 0$ will produce the solution $\rm
f_\pm = \beta_{\pm} =cte$, corresponding to Graham's solutions
obtained in 1993 \cite{Graham2,HDO}). With this ansatz for the
function $\rm s$, we can built $\rm [(\nabla s)^2 \mp\square s]$
term, which differs to $\rm [(\nabla S)^2 \mp\square S]$ by
\begin{equation}
\rm (\nabla s)^2 \mp\square s=(\nabla S)^2 \mp\square S \mp 2
\eta^{\mu \alpha} m_\alpha \frac{\partial S}{\partial q^{\mu}} +
m^\mu m_\mu ,
\end{equation}
for $\rm 2m^\mu \frac{\partial S}{\partial q^{\mu}} \mp m^\mu m_\mu
=0$, we have two cases depending if the constant c is taken in
account or not.
\begin{enumerate}
\item{} for $\rm c\not=0$ and using the superpotential (\ref{super-po})

In this case, $\rm 2m^\mu \frac{\partial S}{\partial q^{\mu}} \mp
m^\mu m_\mu =0$ gives the following equation
\begin{equation}
\rm \frac{e^{3\Omega}gh}{\mu} \left[-6m_0 + 6m_1\frac{\eta_1}{b_2}+
6m_2\frac{\eta_2}{b_3}\right]-2c(-m_0b_1+m_1 b_2 + m_2
b_3)+m_0^2-m_1^2-m2^2=0,
\end{equation}
with solution in the vector $\rm m_\mu=(2cb_1,2cb_2,2cb_3)$ which satisfy
the relation $\rm b_1=\eta_1 +\eta_2$ as defined before.
\item{} for $\rm c=0$ and using the superpotential (\ref{sp})

For this case, is necessary to separate in two independent equations
\begin{eqnarray}
\rm m^\mu m_\mu  &=&0,\label{nula} \\
 \rm \eta^{\mu \alpha}m_\alpha \frac{\partial S}{\partial q^{\mu}} &=&0,
\label{conula}
\end{eqnarray}
where (\ref{nula}) implies that $\rm m_\mu$ is a vector of null
measure (i.e. $\rm -m_0^2 +m_1^2 + m_2^2=0$), and (\ref{conula})
\begin{equation}
\rm \frac{\partial S}{\partial \Omega}m_0 = \frac{\partial
S}{\partial \phi}m_1 + \frac{\partial S}{\partial \sigma}m_2.
\end{equation}
one possibility for the vector $\rm m_\mu$ is the triangle $\rm
\pm(5,3,4)$ and all similarity triangles to this.

When we use the superpotential function S (\ref{sp})
we obtain that the functions $\rm g(\phi)$ and $\rm h(\sigma)$ have
the mathematical structure
\begin{equation}
\rm g(\phi)=g_0 \, e^{\epsilon_1 \Delta \phi}, \qquad h(\sigma)=h_0
\, e^{\epsilon_2 \Delta \sigma}, \label{constra}
\end{equation}
where the constants $\rm \epsilon_1=\frac{3m_0 n_1}{m_2}$ and $\rm
\epsilon_2=\frac{3m_0 n_2}{m_1}$, where $\rm n_i$ satisfy the rule
$\rm n_1+n_2=1$; So, Supersymmetric quantum mechanics constraints
the family of potential fields in the inflation phenomenon to
exponential functions, which corresponds to the third line in the
table (\ref{t:solutions}),  as it has been mentioned in other works
in the literature for this scenario \cite{sodo}.
\end{enumerate}
In the case that both aforementioned equations have no null solution,
the solution for the function $\rm f_\pm$ has the structure
\begin{equation}
\rm f_\pm = b_\pm e^{m_\alpha q^{\alpha}},
\end{equation}
thus, the functions ${\cal B}_\mu$ and ${\cal C}^\nu$ become as
\begin{equation}
\rm {\cal B}_\mu=b_+\, m_\mu e^{m_\mu q^{\mu}} e^{\frac{S}{\hbar}},
\qquad {\cal C}^\mu=\eta^{\mu \nu} b_-\, m_\nu e^{m_\alpha
q^{\alpha}} e^{-\frac{S}{\hbar}}, \label{cb}
\end{equation}

This method was used to obtain the SUSY quantum solution for all
Bianchi Class A models \cite{radames}.

Using the expression for the superpotential function (\ref{sp}) we
see that the only form of S in which these equations are fulfilled,
is when the functions g and h have exponential behaviour. In
\cite{Graham3}, Graham and Luckock mention that the sector
${\cal A}_\pm$ is also distinguished by the existence of a Nicolai
map and a related statistical interpretation of the wave function,
it is say that the Nicolai map in the Grassmann representation only exist in the independent and
fulfilled sectors of the wave function, but not in any other sector.

In a supersymmetric fashion, the calculation by means of the
Grassmann variables  of $|\Psi|^2$  given by (\ref{wavefun}) is well
known  \cite{Faddeev}
\begin{equation}
(\Psi_1|\Psi_2) = \int{ \left( \Psi_1(\theta^*)\right)^*
\Psi_2(\theta^*) e^{ -\sum_i \theta_i^* \theta_i^{} }}  \prod_i
d\theta_i^* d\theta_i^{}, \label{density}
\end{equation}
where the operation  *  is defined as
$(C\theta_1^{}...\theta_n^{})^*=\theta_n^*...\theta_1^*C^*$, with
the usual algebra for the Grassmann numbers $\theta_i \, \theta_j =
- \theta_j \, \theta_i $. The rules to integrate over these numbers
are the following
\begin{equation}
\int{\theta_1^{}\theta_1^*...\theta_n^{}\theta_n^*}d\theta_n^*d\theta_n^{}...
d\theta_1^*d\theta_1^{}=1 \label{grassman3}
\end{equation}
\begin{equation}
\int d\theta_i^* = \int d\theta_i^{} = 0 . \label{grassman4}
\end{equation}

In our case, we have $\Psi_1=\Psi_2=\Psi$. So, when we integrate to
the Grassmann numbers, and employing the relations
 (\ref{grassman3}) and (\ref{grassman4}), we obtain
\begin{equation}
|\Psi|^2 = \bar {\cal A}_+ \, {\cal A}_+^{} + \bar {\cal A}_- \,
{\cal A}_-^{}+ \bar {\cal B}_0 \, {\cal B}_0^{} + \bar {\cal B}_1 \,
{\cal B}_1^{} +\bar {\cal B}_2 \, {\cal B}_2^{} + \bar {\cal C}^0 \,
{\cal C}^0+\bar {\cal C}^1 \, {\cal C}^1 + \bar {\cal C}^2 \, {\cal
C}^2 ,
\end{equation}
where the $\bar {\cal A}$ symbol means the complex operation.

Using the expressions for the functions  ${\cal A}_\pm$, ${\cal
B}_\mu$ and ${\cal C}_\mu$ given in (\ref{As}) and (\ref{cb}), we
arrive to the following expression for the  probability density
\begin{equation}
\rm |\Psi|^2 =  \left[{a_0}_+^2\,+4b_+^2 m_0^2 e^{2(m_0\Omega +m_1
\varphi + m_2 \varsigma)}\right] e^{ \frac{2}{\hbar} S} +\left[
{a_0}_-^2\, +4b_-^2 m_0^2 e^{2(m_0\Omega +m_1 \varphi + m_2
\varsigma)}\right] e^{ -\frac{2}{\hbar} S}. \label{proba}
\end{equation}
Thus, we are able to express (\ref{density}) for our particular
problem.
\section{Conclusions}

Under canonical quantization the Multi-scalar field cosmology of the
anisotropic Bianchi type I model allowed us to determine a family of
potentials that are the most suited to model the inflation
phenomenon. The exact quantum solutions to the Wheeler-DeWitt
equation were found using the Bohmian scheme \citep{bohm} of quantum
mechanics where the ansatz to the wave function  $\rm \Psi(\ell^\mu)
=e^{\frac{a_1}{\hbar} \beta_+ + i\frac{a_i}{\hbar} \beta_-}
W(\ell^\mu) e^{- \frac{S(\ell^\mu)}{\hbar}}$ includes the
superpotential function which plays an important role in solving the
Hamilton-Jacobi equation. The tools of SUSY Quantum Mechanics is
used as an alternative method to test the obtained family of
potentials for the inflation era, such tools restricted the
potentials even further and only to an exponential behavior.
 This method was also used to obtain the SUSY quantum solution for all Bianchi class A Models \cite{radames}.
 Also this class of
solutions appears in the excellent books by Moniz \cite{moniz},
where the author present the review of solutions in quantum and
supersymmetric cosmology for some cosmological models, including the
Bianchi Class A cosmological models, until 2009 year.

\noindent This work was partially supported by CONACYT  167335,
179881 grants. PROMEP grants UGTO-CA-3. This work is part of the
collaboration within the Instituto Avanzado de Cosmolog\'{\i}a. Many
calculations where done by Symbolic Program REDUCE 3.8.



\begin{thebibliography}{99}
\bibitem[J.R.L. Santos \& P.H.R.S. Moraes, 2015]{santos} J.R.L. Santos and P.H.R.S. Moraes
     {\it Fast-roll Solutions from two scalar field inflation}
     \emph{}
     (2015) [arXiv:1504.07204 (gr-qc)].
\bibitem[D. S\'{a}ez-G\'{o}mez, 2008]{gomez} D. S\'{a}ez-G\'{o}mez
     {\it Scalar-Tensor theories and current Cosmology}
     \emph{Problems of Modern Cosmology}
     (2008) [arXiv:0812.1980 (hep-th)].
\bibitem[G. Calcagni \& Andrew R. Liddle, 2007]{andrewl} G. Calcagni and Andrew R. Liddle
     {\it Stability of multi-field cosmological solutions}
     \emph{Phys. Rev. D}
     (2007) [arXiv:0711.3360 (astro-ph)].
\bibitem[M. Capone, C. Rubano, P. Scudellaro, 2006]{capone} M. Capone, C. Rubano and P. Scudellaro
     {\it Slow rolling, inflation and quintessence}
     \emph{Europhys.Lett}
     {\bf 73} 149-155, (2006) [arXiv:astro-ph/0607556].
\bibitem[Juan M.\& J. Socorro, 2013]{juanm} Juan M. Ram\'{i}rez and J. Socorro
     {\it FRW in Cosmological Self-creation Theory}
     \emph{Int. J. Theor. Phys.}
     {\bf 52} 2867-2878, (2013) [arXiv:1206.5413 (gr-qc)].
\bibitem[Copeland et al., 1998]{copeland2} E.J. Copeland, Liddle and D. Wands
     {\it Exponential potentials and cosmological scaling solutions}
     \emph{Phys. Rev. D}
     {\bf 57} 4686, (1998) [arXiv:gr-qc/9711068].
\bibitem[Copeland et al., 2000]{copeland3} E.J. Copeland, T. Barreiro and N.J. Nunes
     {\it Quintessence arising from exponential potentials}
     \emph{Phys. Rev. D}
     {\bf 61} 127301, (2000) [arXiv:astro-ph/9910214].

\bibitem[R. Lazkoz et al., 2007]{quiros} R. Lazkoz, G. León and I. Quiros
     {\it Quintom cosmologies with arbitrary potentials}
     \emph{Phys. Lett. B}
     {\bf 649} 103, (2007) [arXiv:astro-ph/0701353].

\bibitem[M.C. Bento et al., 2001]{bento} M.C. Bento, O. Bertolami and N.C. Santos
     {\it A Two-Field Quintessence Model}
     \emph{Phys. Rev. D}
     {\bf 65} 067301, (2001) [arXiv:astro-ph/0106405].

\bibitem[A.A. Coley \& R.J. van den Hoogen, 2000]{coley} A.A. Coley and R.J. van den Hoogen
     {\it The Dynamics of Multi-Scalar Field Cosmological Models and Assisted Inflation}
     \emph{Phys. Rev. D}
     {\bf 62} 023517, (2000) [arXiv:gr-qc/9911075].


\bibitem[Liddle \& Scherrer, 1998]{andrew}    A.R. Liddle,  and R.J. Scherrer
     {\it Classification of scalar field potential with cosmological scaling solutions}
     \emph{Phys. Rev. D}
     {\bf 59}, 023509 (1998).
\bibitem[Ferreira \& Joyce, 1998]{ferreira}  P.G. Ferreira \& M. Joyce
    {\it Cosmology with a primordial scaling field},
    \emph{Phys. Rev. D},
    {\bf 58}, 023503 (1998).
\bibitem[Copeland et al., 2006]{copeland} E.J. Copeland, M. Sami and S. Tsujikawa
     {\it Dynamics of dark energy}
     \emph{Int. J. Mod. Phys. D}
     {\bf 15} 1753, (2006) [arXiv:hep-th 0603057].

\bibitem[Guzm\'an et al., 2007]{wssa} W. Guzm\'an, M. Sabido, J. Socorro and L. Arturo Ure\~na-L\'opez
    {\it Scalar potentials out of canonical quantum cosmology}
    \emph{Int. J. Mod.  Phys. D }
    {\bf 16} (4), 641-653 (2007).
\bibitem[J. Socorro et al., 2010]{sodo} J. Socorro and Marco D'oleire
    {\it Inflation from supersymmetric  quantum cosmology}
    \emph{Phys. Rev. D}
    {\bf 82}(4), 044008 (2010).
\bibitem[Bohm, 1952]{bohm} D. Bohm
    {\it Suggested interpretation of the quantum theory in terms of  "Hidden" variables I}
    \emph{Phys. Rev.}
    {\bf 85} (2), 166 (1952).

\bibitem[Gibbons \& Gishchuk, 1989]{Gibbons} G.W. Gibbons and L. P. Grishchuk
    {\it Nucl. Phys. B}
    {\bf 313}, 736 (1989).
\bibitem[Zhi, 1987]{Zhi}  Li Zhi Fang and Remo Ruffini, Editors,
    {\it Quantum Cosmology,  Advances Series in Astrophysics and
    Cosmology Vol. 3}
    (World Scientific, Singapore, 1987).
\bibitem[Hartle \& Hawking, 1983]{HH} J. Hartle,  \& S.W. Hawking
    \emph{Phys. Rev. D},
    {\bf 28}, 2960 (1983).
\bibitem[Hawking, 1984]{H} S.W. Hawking
    \emph{ Nucl. Phys. B}
    {\bf 239}, 257 (1984).
\bibitem[Kodama, 1988]{kodama1} H. Kodama
    \emph{Progress of Theor. Phys.}
    {\bf 80}, 1024 (1988).
\bibitem[Kodama, 1990]{kodama2} H. Kodama
    \emph{Phys. Rev D}
    {\bf 42}, 2548 (1990).
\bibitem[Ashtekar, 1989]{ashtekar} A. Ashtekar
    \emph{Phys. Rev. D}
    {\bf 36},1587 (1989).
\bibitem[Moncrief \& Ryan, 1991]{mory} V. Moncrief and M.P. Ryan
    \emph{Phys. Rev. D}
    {\bf 44}, 2375 (1991).
\bibitem[Obreg\'on \& J. Socorro, 1995]{obso} O. Obreg\'on and J. Socorro
    {\it $\Psi=W e^{\pm \Phi}$ quantum cosmological solutions for Class A Bianchi models}
    \emph{Int. J. of Theor. Phys.}
    {\bf 35} (7), 1381 (1995).
\bibitem{witten} E. Witten, Nucl. Phys. B {\bf 188}, 513 (1981).
\bibitem{tkach} V.I. Tkach, J.J. Rosales and O. Obreg\'on, Class. Quantum Grav. {\bf 13},
     2349 (1996).
\bibitem{donets} E.E. Donets, M. N. Tentyukov, M. M. Tsulaia, Phys. Rev. D {\bf 59},
    023515 (1999).
\bibitem{Graham2} R. Graham, Phys. Rev. D {\bf 48}, 1602 (1993).
\bibitem{HDO}P.D. D'Eath, S.W. Hawking and O. Obreg\'on, Phys. Lett. B {\bf 300}, 44 (1993).
\bibitem{radames} J. Socorro and E.R. Medina, Phys. Rev. D {\bf 61}, 087702 (2000).
\bibitem{Graham3} R. Graham and H. Luckock, Phys. Rev. D {\bf 49}, 2786 (1994).
\bibitem{Faddeev}L.D. Faddeev  and  A.A. Slavnov,  {\it Gauge Fields: An
    Introduction to Quantum Theory} (Addison-Wesley, Reading, MA.),
    sec. 2.5. (1991).
\bibitem{moniz} P. Moniz, \emph{ Quantum Cosmology: The supersymmetric
    perspective},  Vol 1 and 2, Lecture Notes in Physics 803 and 804, Springer (2010).


\end{thebibliography}
\end{document}